\documentclass[aps,prb,preprint,superscriptaddress,floatfix,nofootinbib]{revtex4-2}

\usepackage[margin=4cm]{geometry}
\usepackage{amsmath}
\usepackage{xcolor}
\usepackage{placeins}
\usepackage{graphicx,mathrsfs}
\graphicspath{ {figures/}}

\usepackage[colorlinks=true,citecolor=blue,linkcolor=blue,urlcolor=blue]{hyperref}
\usepackage[capitalise]{cleveref}

\begin{document}

\title{Multiphoton absorption and Rabi oscillations in
	 armchair graphene nanoribbons}

\author{B.~S.~Monozon}
\affiliation{Physics Department, Marine Technical University, 3 Lotsmanskaya Str.,
	190121 St.Petersburg, Russia}
\author{P.~Schmelcher}
\affiliation{Zentrum f\"ur Optische Quantentechnologien, Universit\"at
	Hamburg, Luruper Chaussee 149, 22761 Hamburg, Germany}
\affiliation{The Hamburg Centre for Ultrafast Imaging, Universit\"at
	Hamburg, Luruper Chaussee 149, 22761 Hamburg, Germany}

\begin{abstract}
We present an analytical approach
to the problem of the multiphoton absorption
and Rabi oscillations in an armchair
graphene nanoribbon (AGNR)
in the presence of a time-oscillating strong electric field
induced by a light wave
directed parallel to the ribbon axis.
The two-dimensional Dirac equation for the massless
electron subject to the ribbon confinement is employed.
In the resonant approximation the electron-hole
pair production rate, associated with the
electron transitions
between the valence and conduction size-quantized subbands,
the corresponding
multiphoton absorption coefficient and the
frequency of the Rabi oscillations are obtained
in an explicit form. We trace the dependencies
of the above quantities on the ribbon width
and electric field strength for both the multiphoton
assisted and tunneling regimes relevant to
the time-oscillating and practically constant
electric field, respectively. A significant
enhancement effect
of the oscillating character of the electric field
on the intersubband transitions
is encountered.
Our analytical results are in qualitative agreement
with those obtained for the graphene layer by numerical methods.
Estimates of the expected experimental values
for the typically employed AGNR and laser parameters
show that both the Rabi oscillations and multiphoton absorption
are accessible in the laboratory.
The data relevant to the intersubband tunneling
makes the AGNR a 1D condensed matter analog in
which the quantum electrodynamic vacuum decay
can be detected by applying
an external laboratory electric field.
\end{abstract}

\maketitle

\section{Introduction}\label{S:intro}

The pioneering work by Wallace \cite{ Wall}
devoted to the electron states in graphene, an
ideal 2D crystal of carbon atoms arranged on a honeycomb
lattice, has lead to a series of experimental
and theoretical investigations resulting in a vast
literature
(see Refs. \cite{Cast1, Das} and references therein).
This undiminished interest is caused by its unique
mechanical, electronic, optical and transport properties
\cite{Cast2}, that in turn is the consequence of the graphene
electron nature. In the vicinity of the $\vec{K}^{\pm}$
corners of the graphene Brillouin zone
(corresponding to Dirac points)
the electron dispersion law looks like
$\varepsilon = \hbar v_{\tiny{F}}k$, where
$\varepsilon~\mbox{and}~\vec{k}$
are the energy and 2D wave vector, respectively,
counted from the Dirac points, where
$v_{\tiny{F}}=10^6~\mbox{m/s}$ is the graphene Fermi velocity.
This linear law results in the low energy
electron states to be governed by a massless 2D Dirac equation
\cite{ Wall, Cast1}. The numerous branches
of graphene studies might be grouped into two
classes. The first one are investigations
contributing to corresponding applications
(phase modulators, nanopore sensors, ballistic transistors
photodetectors, topological insulators
(see Ref. \cite{ Fill}) and
fundamental science (Klein tunneling, Zitterbewegung,
quantum Hall effect), both related to condensed
matter physics. In the second class of works
graphene plays the role of a physical
environment, in which some of the basic
quantum electrodynamic (QED) effects, predicted earlier,
can be verified in a laboratory. The existing
experimental facilities prevent these effects
to be observed in vacuum. The linear dispersion law
for the massless fermions up to the energy
$\varepsilon \simeq 1~\mbox{eV}$ allow us
to consider
graphene as a condensed matter counterpart
for relativistic quantum field theory and
a filled Fermi sea in graphene as a filled Dirac sea
in QED vacuum \cite{All}.

The vacuum decay in the presence of a strong
classical electromagnetic field, accompanied
by the electron-positron pair production (pp),
remains one of the most important QED effects.
Originally, it has been predicted by Sauter \cite{Saut}
in the context of the Klein paradox \cite{Klein},
firmly theoretically described by Schwinger
\cite{Schwin}, commented by Nikishov
\cite{Nik} and Cohen and McGady \cite{Coh} and
investigated by the $S$-matrix \cite{NarNik,Nik1}
and quasi-classical \cite{BrezIt,Pop} methods
(see Ref. \cite{Fedot} for details)
to give for the pp rate

\begin{equation}\label{E:ratevac}
W^{v}\sim\exp \left(-\frac{\pi
F_{c}^{v}}{F}\right).
\end{equation}
In equation (\ref{E:ratevac})
$F~\mbox{and}~F_{c}^{v}=
\frac{m^2 c^3}{e\hbar}$
are the time-independent external and critical
electric fields, respectively.
The latter provides the balance between the energy an
electron acquires for the Compton wavelength and a gap $2mc^2$
in the vacuum energy spectrum. The experimental
observation of the Schwinger vacuum decay is challenged
by the extremely strong critical field
$F_{c}^{v}\simeq 10^{13}~\mbox{kV/cm}$,
which considerably exceeds the currently highest possible
experimental values
$F_{\mbox{\tiny{exp}}}\simeq 10^{-2}
F_{c}^{v}$
\cite{Akal19}, which leads to the
exponential suppression of the pp rate $W^{v}$.

In the case of a time-oscillating electric field of
magnitude $F_0$ and frequency $\omega$, the
particle-field interaction in vacuum is determined by the Keldysh
parameter \cite{Keld65}

\begin{equation}\label{E:keld1}
\gamma^{\tiny{(v)}}=\frac{\omega m c}{e F_0},
\end{equation}
distinguishing the Schwinger interband
tunneling $(\gamma^{\tiny{(v)}}\ll 1)$ and multiphoton
assisted $(\gamma^{\tiny{(v)}} \geq 1)$
mechanisms of the pp.
The exact equations for the numbers of
fermion and boson pairs, created from the
vacuum by the time-periodic electric field,
have been derived by Mostepanenko and Frolov \cite{Most}.
The enhancement
of the pp output by the high frequency electric field
was discussed in the works
\cite{Dun, Coop, Klug, Bro, Bord, Klim}.
It is noteworthy that the multiphoton pp from the vacuum
becomes insignificant for frequencies less
than those in the gamma region \cite{KlimM, Grib}.
The decay of the arbitrary dimensional
QED, induced by the quasi-constant electric field,
has been studied by Gavrilov and Gitman
in Ref. \cite{GavGit}, in which,
in particular, the mean numbers of created
bosons and fermions particles were determined.
Recently Taya \cite{Taya}, based on the Dirac
equation and Furry approach \cite{Furry},
comprehensively studied and thoroughly
reviewed the problem of the interplay between
the low frequency strong and fast oscillating
perturbative electromagnetic fields. It was shown
that such field interaction results in the
significant growth of the electron-positron
production from vacuum.
In \cite{Taya21} the vacuum pp by a time-dependent strong
electric field on the basis a semiclassical Wentzel-
Kramers-Brillouin analysis has been explored.

In order to gain insight into the mechanisms
of the vacuum decay accompanying by the electron-positron
pp, the methods, relevant to the 3+1-dimensional
QED vacuum, have been extended to the similar
semiconductor structures and to the vacuum 2+1 analog,
namely graphene layer, related to the electron-hole (e-h) pp.
Linder et al \cite{Lind} employed the parallel
between the Dirac and two-band equations for the
electron-positron in vacuum and e-h pairs in the
narrow-gapped semiconductors \cite{Keld64}, respectively.
The enhancement of the pair creation Landau-Zener
probability \cite{Land, Zin}, caused by the weak
time-oscillating electric field superimposed on
the strong constant \cite{Schwin} or slowly varying
Sauter pulse \cite{Saut} electric fields, has been discussed.
Thus, the narrow-gapped semiconductors become the media
for the laboratory testing of the fundamental
quantum field theory predictions.

For a constant electric field the pp rate has been
calculated both analytically and numerically
for the gapless (massless) and gapped (massive)
graphene \cite{All, KlimM, Akal16, GavGitYok}.
In particular, Allor et.al. \cite{All}
proposed an experimental test
for the Schwinger tunneling mechanism.
Effects of the time-dependent electric field
on the pair creation in the gapped and gapless
graphene have been investigated for periodically oscillating
\cite{Avet, Akal16}, different time pulses
\cite{Fill1, Fill}, and Sauter-like \cite{Saut, KlimM}
electric fields. Recently, Akal \emph{et al} \cite{Akal19}
studied dynamically-assisted pp in gapped
graphene monolayers subject to bichromatic electric fields.
Gagnon \emph{et al} \cite{Gag}
investigated the dynamical pair creation in the presence
of a strong magnetic fields directed perpendicular
to a graphene monolayer.

As it follows from the theoretical results of the works
listed above, gapped graphene is suitable
as a condensed matter emulator
for the experimental test of the pp
in vacuum, induced by the electric field. The point is that
the electric fields, needed for the
particle-antiparticle pair creation in graphene, can be realized
in modern experimental labotatories in contrast
to vacuum, for which the corresponding fields
remain unattainable. The gapped graphene critical
electric field $F_{c}^{g}$
can be obtained from $F_{c}^{v}$
in eq. (\ref{E:ratevac}) by replacing $c~\mbox{by}~v_F$
and $m$ by
$\Delta_{g}/v_{\tiny{F}}^2~\mbox{where}~\Delta_{g}$
is a bandgap, induced, for example, by epitaxial growth
on a suitable substrate \cite{Zho}. For the realistic
bandgap $\Delta_{g}\simeq 0.3~\mbox{eV}$ \cite{Zho}
the critical electric field is
$F_{c}^{g}\simeq 1.6\cdot10^3~\mbox{kV/cm}$,
that is much less than the vacuum field
$F_{c}^{v}$.

For the time-oscillating
electric field the Keldysh parameter $\gamma^{\tiny{(g)}}$
is obtained from $\gamma^{\tiny{(v)}}$
in eq. (\ref{E:keld1}) by the same replacements
as those associated with the field $F_{c}^{g}$.
For the infrared electromagnetic wave
$(\hbar \omega \simeq 0.1 ~\mbox{eV})$,
providing the multiphoton mechanism
of the pair creation
$(\gamma^{\tiny{(v)}}=\gamma^{\tiny{(g)}}\simeq 1,~
\Delta_{g}\simeq 0.3~\mbox{eV})$, the
field magnitudes for the vacuum and gapped graphene
become $F_{0}^{v}\simeq 3\cdot10^{6}~\mbox{kV/cm}
~\mbox{and}~F_{0}^{g}\simeq 5\cdot10^2~\mbox
{kV/cm}$,
respectively. The graphene related electric fields correspond
to the laser intensity $I\simeq 6.5\cdot10^5~\mbox {kW/cm}^2$,
which can be comfortably reached by present laser technology.
In gapless graphene $(\Delta=0)$ the zero effective mass
$m=0$ ensures the pp for any electric fields and
frequencies \cite{Avet} and prevents the
existence of a critical electric field and exponential
suppression. Since the latter is the basic signature
of the Schwinger vacuum tunneling, the intrinsic gapless
2D graphene layer does not seem to be an ideal
candidate for examining QED phenomena.

At this stage another gapped graphene structure is demanded
to simulate the electric field induced pp production from vacuum.
Armchair graphene nanoribbon (AGNR)-a quasi-1D graphene
strips with width $d$, surpasses in some sense the graphene layer.
In graphene the bandgap $\Delta_{g}$
is the parameter of the theory, associated with the
technical parameters
(epitaxial growth, elastic strain, Rashba spin splitting
on magnetic substraits \cite{Akal16}),
modifying the genuine graphene properties,
while the AGNR gap $\Delta_{r}\sim d^{-1}$ \cite{BrFert}
is the intrinsic parameter of the untouched graphene.
In the case of necessity the bandgaps can be measured
experimentally, in particular, by optical methods.
Since both the AGNR and gapped graphene are
semiconductor-like structures, the interband optical
absorption spectrum in the vicinity of the threshold
$\hbar\omega = \Delta$ demonstrates an easily detected
singularity
$\left(\hbar\omega - \Delta_{r}  \right)^{-\frac{1}{2}}$
and a weakly manifested step-like form
$\Theta (\hbar\omega - \Delta_{g})$
for the quasi-1D AGNR and 2D graphene layer, respectively.
Clearly,
the first is favorable for the precise bandgap measurement.
The one-photon absorption of the low intensity
light in AGNR has been theoretically studied in a broad range
of works \cite{Bar, Yang, Hsu,
Prez, Gun, Ingl},
based on various computational techniques,
as well as those related to the
graphene layer \cite{Fill}, \cite{Akal19}, \cite{Akal16},
\cite{Avet, Fill1, Gag}.
Undoubtedly, numerical approaches are preferable
for an adequate description of concrete experiments.
Nevertheless, analytical approaches,
being the focus of the present work,
are indispensable
to elucidate the basic physics of AGNR by deriving
closed form analytical expressions for their
properties. Our second goal is to promote the application
of the AGNR based materials in high-power opto-electronics,
using the transparent dependencies
of their underlying effects on the
ribbon width and strong light wave intensity.
However, to the best of our knowledge,
in contrast to the graphene layer,
explicit results relevant
to the dynamically assisted e-h production and
accompanying multiphoton optical absorption in
AGNR, irradiated by the intensive light, have
virtually not been addressed in the literature yet.

In order to fill this gap
we analytically determine the production rate
of the e-h pairs, derive the multiphoton
interband absorption coefficient and Rabi oscillations frequency
in AGNR subject to a time-oscillating electric field
of a strong light wave. In addition,
we present the pp rate, corresponding to the interband
tunnelling, caused by the time-constant electric field.
We exploit the nearest-neighbors
tight-binding model, generating the Dirac
equation, to govern the low-energies
graphene fermions \cite{Wall}, \cite{Cast1} and resonant
approximation for particle-wave interaction.
The presented results can be employed in two ways.
First, properties of the interband multiphoton absorption and
e-h tunneling render AGNR
a unique material, contributing significantly to
applied and fundamental fields of solid state
physics. Second, these properties of the graphene ribbon,
interpreted as a quantum field theory object,
qualitatively highlight the electrically induced instability
of a QED vacuum.

The paper is organized as follows. In Sec. II the general
approach and basic analytical equations are presented.
The Rabi oscillation frequency, rates of the intersubband
transitions, associated with the tunneling and multiphoton assisted
mechanisms, are derived,
discussed and estimated with respect to future experiments
in Sec. III. Sec. IV contains our conclusions.

\section{General approach}\label{S:gen}

We consider an electrically biased AGNR,
with width $d$
and length $L$, placed on the $x-y$ plane and bounded by
straight lines $x=\pm d/2.$
The time-oscillating electric field
$F(t)=F_{0}\cos \omega t$,
with magnitude $F_{0}$ and frequency $\omega$,
as well as the polarization of
the light wave,
are chosen to be parallel to the ribbon $y$-axis.
The energy spectrum of the free electron in AGNR,
derived by Brey and Fertig \cite{BrFert},
is a sequence of the 1D subbands with
the energies

\begin{eqnarray}\label{E:subb}
\pm E_N(k)&=& \left(\varepsilon_N^2 + \hbar^2 v_{\tiny{F}}^2 k^2
\right)^{\frac{1}{2}};\quad
\varepsilon_N=|N-\tilde{\sigma}|\frac{\pi\hbar v_{\tiny{F}}}{d};
\nonumber\\
~N&=&0,\pm1,\pm2,\ldots~;
\quad\tilde{\sigma}=\frac{Kd}{\pi}-\left[ \frac{Kd}{\pi}\right],
\end{eqnarray}
where $\varepsilon_N~\mbox{and}~\hbar v_{\tiny{F}} k$
($k$ is the wave number) are the size-quantized and continuous
energies, corresponding to the transverse $(x)$ and longitudinal $(y)$
motions, respectively.
$v_{\tiny{F}} = 10^6~\mbox{m/s}~\mbox{and}~
K = 4\pi / 3a_0 ~(a_0 = 2.46\, \mbox{{\AA}}
~\mbox{is the graphene lattice constant})$ are the
Fermi velocity in graphene and the wave number,
determining the nonequivalent Dirac points
$\vec{K}^{(+,-)}= (\pm K, 0)$
in the graphene Brillouin zone.
Below, to be specific, we will consider AGNR
of the family $\tilde{\sigma} = 1/3$, providing
a semiconductor-like gapped structure.

The general approach to the problem of the optical
absorption in AGNR, associated with the interband
electron transitions, has been developed in Ref. \cite{Sas}.
For a $y$-polarized light wave transitions are allowed
between the valence and conduction bands, possessing the energies
$\mp\mid E_N (k)\mid$,
respectively. In \cite{Sas}
the dynamical conductivity has been chosen
to describe the graphene
optical properties. Here, however, we take the
traditional for the semiconductor-like structures equivalent
characteristic, namely the $l$-photon absorption coefficient
$\alpha^{(l)}$, linked to the transition probability
$W^{(l)}$ by the following relation

\begin{equation}\label{E:coeff}
\alpha^{(l)} =
\frac{\hbar \omega}{n_b \varepsilon_0 c S F_{0}^2}W^{(l)};\quad
W^{(l)} = \sum_N W_{N}^{(l)}.
\end{equation}
In this equation $n_b$ is the refractive index of the ribbon substrate,
$c$ is the speed of light, $S=Ld$ is the area of the ribbon,
and $W_N^{(l)}$ is the probability of the transition
between the valence and conduction $N$-subbands
per unit length per unit time, i.e., the length density
of the e-h pair production  rate.

The equation, describing the electron at a position
$\vec{r}(x,y)$ subject to the external time-dependent
electric field, possesses the form of a Dirac equation

\begin{equation}\label{E:basic}
\hat{{ H}}\vec{\Psi}(\vec{r},t)={{i}}\hbar\frac{\partial
\vec{\Psi}(\vec{r},t) }{\partial t},
\end{equation}
where the Hamiltonian

\begin{equation}\label{E:hamilt}
\hat{{ H}}(\vec{k};y,t) = \hat{{ H}}_x (\hat{k}_x)
 + \hat{{ H}}_y (\hat{k}_y)
+ \hat{{I}}\left(-eyF(t)\right) ;~\hat{\vec{k}}=-i\vec{\nabla}
\end{equation}
is formed by the Hamiltonians

$$
\hat{{ H}}_j (\hat{k}_j)=
\hbar v_{\tiny{F}}\left(\begin{array}{cc} -\sigma_j \hat{k}_j
& 0 \\0 & \sigma_j^{*} \hat{k}_j \\
\end{array} \right),
$$
relevant to the nonequivalent Dirac points $\vec{K}^{(+,-)}$ \cite{BrFert}
and the electric field potential $-e y F(t)$.
The matrices $\hat{{I}}~\mbox{and}~\vec{\sigma}$
are the unit and Pauli matrices, respectively.

Further we choose the wave function $\vec{\Psi}$,
associated with the $N$ subband, in the form

\begin{equation}\label{E:totfun}
\vec{\Psi}_{N}(\vec{r},t)=
\frac{1}{\sqrt{2}}\left[ u_{NA}(y,t)\vec{\Phi}_{NA}(x) +
u_{NB}(y,t)\vec{\Phi}_{NB}(x)\right],
\end{equation}
where $\vec{\Phi}_{NA(B)}~\mbox{and}~u_{NA(B)}$
are the wave functions describing the
electron transverse $x$- and longitudinal $y$-states,
governed by the ribbon confinement and electric field
$F(t)$, respectively. The indices $A(B)$ mark
the graphene sublattices. The explicit form
and properties of the four component
functions $\vec{\Phi}_{NA(B)}$
are presented in Ref. \cite{MonSchm12}.
The total $x$-wave function

$$
\vec{\Phi}_{N} (x)=\frac{1}{\sqrt{2}}\left [ \vec{\Phi}_{NA} (x) +
\vec{\Phi}_{NB} (x)  \right ]
$$
and the sublattice wave functions $\vec{\Phi}_{NA(B)}$ satisfy the equations

\begin{eqnarray}\label{E:cond}
\begin{array}{l}
\hat{H}_x (\hat{k}_{x})\vec{\Phi}_{NA(B)} =\varepsilon_N \vec{\Phi}_{NB(A)};~\\
\hat{H}_x (\hat{k}_{x})\vec{\Phi}_{N} (x)=\varepsilon_N \vec{\Phi}_{N}(x);\\
\langle\vec{\Phi}_{N'B(A)}|\vec{\Phi}_{NA(B)}\rangle =0~;\\
\langle\vec{\Phi}_{N'A(B)}|\vec{\Phi}_{NA(B)}\rangle =
\langle\vec{\Phi}_{N'}|\vec{\Phi}_{N}\rangle =\delta_{N'N},
\end{array}
\end{eqnarray}
where the size-quantized energy levels $\varepsilon_N$
are given in eq. (\ref{E:subb}).

Substituting the wave function
$\vec{\Psi}_{N}$ (see eq.(\ref{E:totfun}))
into eq. (\ref{E:basic})
and in view of eq. (\ref{E:cond}) we arrive after routine
manipulations to a set of equations
for the functions $u_{NA(B)}$ with the dropped indices $N$

\begin{flalign}\label{E:set1}
&\left[\left(-{i}\hbar\frac{\partial}{\partial t} -eyF(t)\right){\hat{I}}+
\varepsilon_N \hat{\sigma}_x
-{i}\hbar v_{\tiny{F}}\frac{\partial}{\partial y}\hat{\sigma}_y
\right]\vec{u}=0 \\
&~\vec{u}(y,t)= (u_A (y,t) , u_B (y,t) )\nonumber.
\end{flalign}

Solving eq. (\ref{E:set1}) the wave function $\vec{\Psi}_{N}$
in eq. (\ref{E:totfun}) can be calculated in principle.
However, this function does not bring us closer
to the solution of the problem of the multiphoton interband transitions.
The point is that this problem implies two effects
of the electric field $F(t)$, namely the formation
of the valence and conduction intraband time-dependent electron states
and the generation of the interband transitions between them.

In order to highlight the operators, responsible
for these effects, we transform eq. (\ref{E:set1})
by the substitution

\begin{flalign}\label{E:ustate}
& \vec{u}(y,t) =
\frac{1}{\sqrt{2}}(\hat{\sigma}_z  + \hat{\sigma}_x)
\hat{U}^{+}\exp [ {{i}} q(t)y] \vec{\eta}(t), \\
& q(t)= \frac{e}{\hbar}\int_0^t F(\tau) d\tau + k,
~~\vec{\eta}(t)= (\eta_1 (t) , \eta_2 (t)), \nonumber
\end{flalign}
containing the longitudinal momentum $k$. The unitary
operator

\begin{flalign}\label{E:folwout}
&\hat{U}^{+} = \frac{(\Omega_N + \omega_N){\hat{I}} -{i}
v_{\tiny{F}}q(t)\hat{\sigma}_x}
{[2\Omega_N(\Omega_N + \omega_N)]^{\frac{1}{2}}};~ \\
&\Omega_N^2(t)=\omega_N^2 +v_{\tiny{F}}^2
q(t)^2;~\omega_N=\frac{\varepsilon_N}{\hbar} \nonumber
\end{flalign}
provides the equation of motion
\begin{equation}\label{E:separ}
{{i}}\hat{I}\dot{{\vec{\eta}}}=
(\Omega_N (t)\hat{\sigma}_z  - R_N (t) \hat{\sigma}_x)\vec{\eta};~
R_N (t) =
\frac{\omega_N v_{\tiny{F}} \dot{q}(t)}
{2 \Omega_N^2(t)}
\end{equation}
for the wave function $\vec{\eta}(t)$.
The applied transformation splits the electric field
term in eq. (\ref{E:set1}) into two components
$\sim \Omega_N (t)$ and $\sim R_N (t)$,
attributed to the separated valence and conduction intraband states
and interband transitions
between them, respectively. This transformation is
analogous to the Foldy-Wouthuysen one \cite{FolWout},
separating completely for $F(t)= 0$ the intraband
valence and conduction states, corresponding
to the energies $\pm E_N(k)$ (\ref{E:subb}), respectively.
In particular, this method has been employed in
works dedicated to the interband
magneto-electroabsorption \cite{ArPik} and
multiphoton magnetoabsorption \cite{ZhilMon78}
in the narrow-gapped bulk semiconductors.

Substituting the function $\eta_{1,2}$ in the form

$$
\eta_{1,2}(t)= f_{1,2}(t)
\exp \left[ \mp {i}\int_0^t \Omega_N (\tau) d\tau \right]
$$
into eq. (\ref{E:separ}), we arrive to a set of equations

\begin{equation}\label{E:set2}
{i} \dot{f}_{1,2}(t) =
-R_N (t)\exp\left[ \pm 2 {i}\int_0^t \Omega_N (\tau) d\tau
\right]f_{2,1}(t).
\end{equation}
For the case of a periodically oscillating electric field
$F(t) = F_0 \cos \omega t$ we select the periodic part in the
exponential factor in eq. (\ref{E:set2})

\begin{equation}\label{E:period}
\exp\left[ 2 {i}\int_0^t \Omega_N (\tau) d\tau \right]=
\exp\left(\frac{{i}}{\hbar}\mathscr{E}_N t  \right)S_N(t),
\end{equation}
where $S_N(t) =S_N\left(t+ \frac{2\pi}{\omega}\right)$ and

\begin{equation}\label{E:quasi}
\mathscr{E}_N=\frac{\hbar \omega}{\pi}
\int_{-\frac{\pi}{\omega}}^{+\frac{\pi}{\omega}} \Omega_N (t)d t.
\end{equation}
Since the term $\hbar^2 \Omega_N^2(t)$
is the eigenvalue of the intraband Hamiltonian
$(-\hbar v_{\tiny{F}} q(t)\hat{\sigma}_y +\varepsilon_N\hat{\sigma}_z)^2$,
the quasienergy $\mathscr{E}_N$  (\ref{E:quasi})
can be treated as the change of the electron quasienergy, caused
by the transition between the valence and conduction
$N$ subbands or, equivalently, the quasienergy of the created
e-h pair. In addition, it is reasonable
to exploit in eqs. (\ref{E:set2}) the expansion of the
time-periodic product $R_N(t)S_N(t)$ in the Fourier series

\begin{equation}\label{E:expan}
R_N(t)S_N(t)= \sum_{l=-\infty}^{+\infty}A_l{e}^{-{il\omega t}},
\end{equation}
where

\begin{equation}\label{E:four}
A_l(\omega) = \frac{\omega}{2\pi}
\int_{-\frac{\pi}{\omega}}^{+\frac{\pi}{\omega}}R_N(t)S_N(t){e}^{{i}l\omega
t}dt.
\end{equation}

Since the exact equations (\ref{E:set2}) do not admit an
analytical solution, further we consider this set
in the resonant approximation
$\omega_l \ll \omega~(\omega_l =
\mathscr{E}_N / \hbar  -l \omega )$, implying the significant
proximity of the e-h pair quasienergy
$\mathscr{E}_N$ to the total energy $l\hbar \omega$
of $l$ involved photons. Averaging the coefficients
in eqs. (\ref{E:set2}) in view of eqs.
(\ref{E:period}), (\ref{E:quasi}) and (\ref{E:expan}),
we obtain the following relations

\begin{equation}\label{E:aver}
{i}\dot{\bar{f}}_1 =
-\bar{f}_2 A_l {e}^{{i}\omega_l t};\quad
{i}\dot{\bar{f}}_2 =
-\bar{f}_1 A_l^{*} {e}^{-{i}\omega_l t},
\end{equation}
where $\bar{f}_{1,2}(t)$ are the functions
$f_{1,2}(t)$ averaged over the electric field
period $T=2\pi/\omega$ in the vicinity of the
time instant $t$

$$
\bar{f}_{1,2}(t) = \frac{1}{T} \int_{t - \frac{T}{2}}
^{t - \frac{T}{2}}f_{1,2}(\tau) d\tau.
$$

Equations (\ref{E:aver}) describe the
well known two-level problem \cite{LandLif}.
Under the initial conditions
$\bar{f}_{1}(0)=0,~\bar{f}_{2}(0) = 1$
the solution to eqs. (\ref{E:aver}) becomes

\begin{eqnarray}\label{E:sol}
\bar{f}_1(t) &=& {i} A_l
\frac{\sin \lambda_l t}{\lambda_l}
{e}^{\frac{{i}}{2}\omega_l t},
\nonumber\\
\bar{f}_2(t) &=&\left(\cos \lambda_l t+
\frac{{i}\omega_l}{2\lambda_l} \sin \lambda_l t  \right)
{e}^{-\frac{{i}}{2}\omega_l t},
\nonumber\\
\lambda_l & =&  \left(|A_l|^2 +\frac{1}{4}\omega_l^2    \right)^{\frac{1}{2}},~
l=1,2,\ldots ,
\end{eqnarray}
where the quasienergy $\mathscr{E}_N$
and coefficients $A_l$ are given by eqs. (\ref{E:quasi})
and (\ref{E:four}), respectively. The derived equations
are valid under the condition $\lambda_l \ll \omega$,i.e.
the resonant approximation.

From the above procedure of determining the functions
$\bar{f}_{1,2}(t)$ (\ref{E:sol}) and their meanings
the differential probability $w_N^{\tiny (l)} (k)$
of the interband electron transition
between the valence and conduction $N$th subbands
can be written in the form $w_N^{\tiny (l)} (k) = |\bar{f}_{1}(t)  |^2$,
to give in view of eq. (\ref{E:sol})

\begin{equation}\label{E:probab}
w_N^{\tiny (l)} (k) = |A_l(k)|^2 \frac{\sin ^2\lambda_l t}{\lambda_l^2}.
\end{equation}
In an effort to continue the analytical
investigations the explicit form of the coefficient $A_l$
is needed. In view of eqs. (\ref{E:separ})
and (\ref{E:period}) for the functions
$R_N(t)~\mbox{and}~S_N(t)$, respectively,
and under the condition $\mathscr{E}_N = l\hbar \omega$
the coefficient $A_l(k)$ in eq. (\ref{E:four})
acquires the closed integral form

\begin{flalign}\label{E:four1}
&\frac{2\gamma_N}{\omega}A_l(k)=
\frac{1}{2\pi}\int_{-\pi}^{+\pi}
\frac{\exp\left[{i}l
\int_{0}^{\varphi}\Lambda^{\frac{1}{2}}(\psi, u)d\psi\right]}
{\Lambda (\varphi, u)}\cos \varphi d\varphi;
\nonumber\\
&\Lambda (\varphi, u)=1 +\gamma_N^{-2}(\sin \varphi + u)^2,\\
&u=\frac{\hbar \omega k}{eF_0},
\quad \gamma_N = \frac{\omega \Delta_N}{2ev_{\tiny{F}}F_0}, \nonumber
\end{flalign}
while the quasienergy $\mathscr{E}_N$
is obtained explicitly

\begin{flalign}\label{E:enex}
&\mathscr{E}_N (k) = \frac{2}{\pi} \Delta_N
\frac{1}{s_N}E\left(\sqrt{1-s_N^2}\right)
\left(1 + s_N^2 \frac{2\hbar^2 v_{\tiny{F}}^2 k^2 }{\Delta_N^2} \right); \\
&\quad s_N^2 = (1+\gamma_N^{-2} )^{-1}. \nonumber
\end{flalign}
In eqs. (\ref{E:four1}), (\ref{E:enex}) $\Delta_N = 2\varepsilon_N$
is the intersubband energy gap, $E(x)$ is the complete
elliptic integral of the second kind \cite{Abram}
and $\gamma_N$ is the Keldysh adiabaticity  parameter
\cite{Keld65}, determining the intersubband transition
mechanism. For $\gamma_N \ll 1$ the e-h pp is generated
by the Zener tunneling in an approximately constant
electric field $F_{0}$, while for $\gamma_N \gg 1$
the multiphoton assisted mechanism dominates.

The employed further condition $l\gg 1$ allows us
to perform analytically the integration in eq. (\ref{E:four1})
by the steepest-descent method, treating
the number of absorbed photons $l$ as a large parameter.
Note that the below-determined total probability
includes the term
$\sim |A_l(k)|^2 g(k)\frac{dk}{d\mathscr{E}_N}$,
where for the quasi-1D
AGNR the density of $k$-states $g(k)= \mbox{const}$
and $\frac{dk}{d\mathscr{E}_N} \sim k^{-1}$
(see eq. (\ref{E:enex})). Thus, the term $|A_l(0)|$
is sufficient for the description of the optical transition
in the vicinity of the spectral singularity $k=0$.

Since the differential probability
$\sim |A_l(k)|^2$ (\ref{E:probab}) and
the corresponding Rabi oscillation
frequencies $\sim |A_l(k)|$ reach a maximum for $k=0$ both in graphene
\cite{Akal16},
and in the AGNR (see eq. (\ref{E:four1})),
we further focus on the coefficient
$|A_l(0)|$.
Though analogous
integrals have appeared earlier in Refs.
\cite{Keld65}, \cite{Keld58}, \cite{Perel},
a brief
outlook onto the needed techniques
seems to be in place here. The integration segment
$(-\pi,0)\rightarrow +(\pi,0)$ in eq. (\ref{E:four1})
is replaced by a contour, consisting of
additional segments
$(\pi,0)\rightarrow (\pi,-\infty)\rightarrow
 (-\pi,-\infty)\rightarrow (-\pi,0)$.
Note that the saddle points
$\varphi_n = \arcsin ({i}\gamma_N +n\pi);~
n=0,\pm1$, at which the function $\Lambda (\varphi,0)$,
vanishes, coincide with the poles of the integrand.
In this case the steepest-descent method implies
the subsequent bypassing of the saddle points
$\varphi_{1},\varphi_0,\varphi_{-1},$
positioned inside the corresponding three segments.
The bypassing arcs with the vanishing small
radius are $2\pi/3~\mbox{and}~4\pi/3$ for the
saddle points $\varphi_{\pm 1}~\mbox{and}~\varphi_0$,
respectively. In view of the saddle points total
contribution, the calculation reduces to the
contour integration around the pole $\varphi_0$.
The residue theorem gives then

\begin{equation}\label{E:four2}
A_l (0) = \frac{\omega}{3}
\exp \left\{ -\frac{l}{s_N}\left[K(s_N) - E(s_N)  \right] \right\}
\sin^2 \frac{l\pi}{2},
\end{equation}
where $K(x)$ is the complete elliptic integral
of the first kind \cite{Abram}.
The factor $\sin^2 \frac{l\pi}{2}$
stems from the interference of the saddle points
and the condition $l\hbar\omega = \mathscr{E}_N$
(see Ref. \cite{Kov} for more details).
\\
\\
\section{Results and discussion}\label{S:Resdisc}

\subsection{ Rabi oscillations}

\begin{figure}[h]
	\centering
	\includegraphics[width=1\columnwidth]{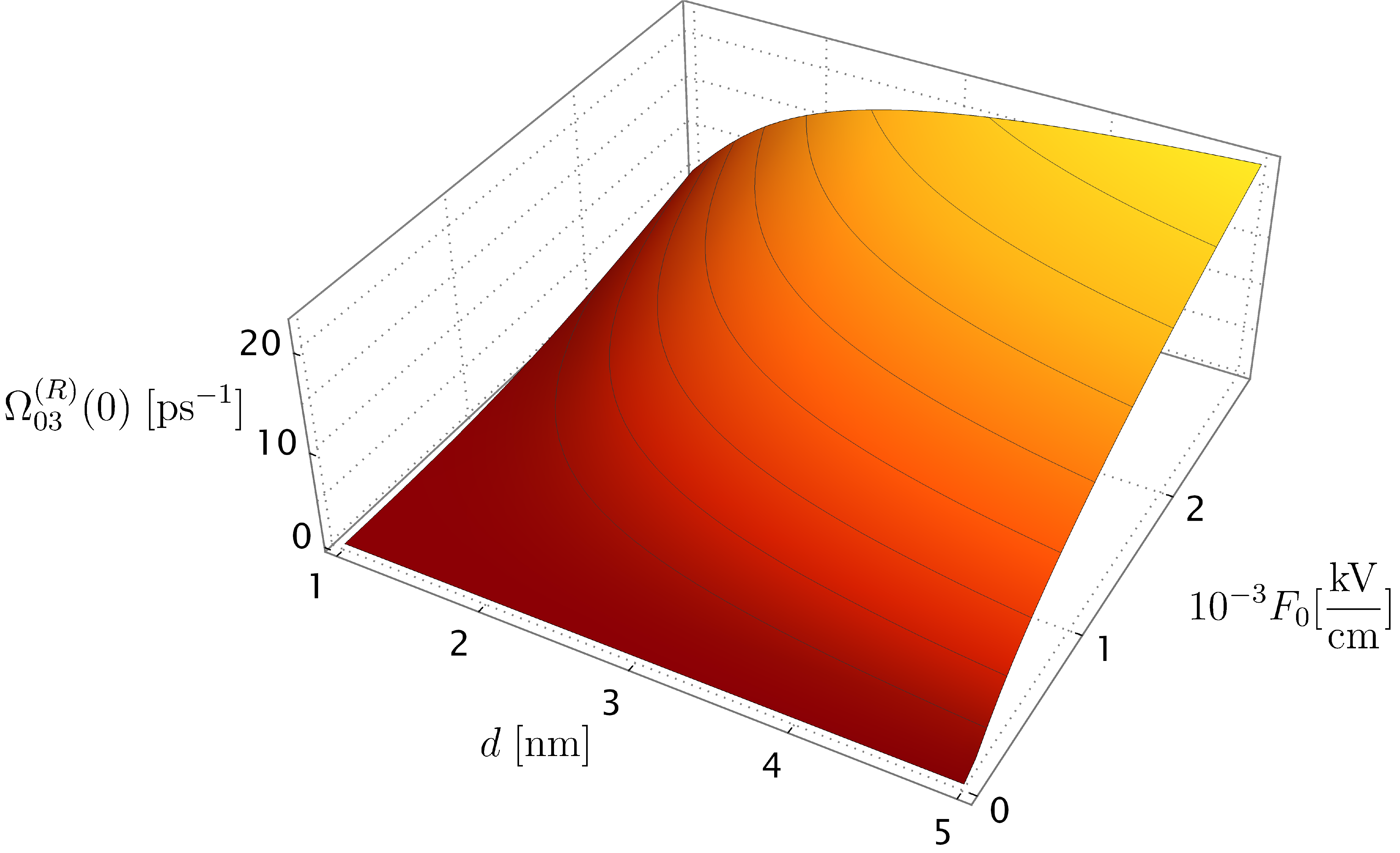}
	\caption{The Rabi frequency $\Omega_{03}^{(\tiny{R})}(0)$
		versus the ribbon width $d$ and electric field
		$F_{0}$. The frequency
		$\Omega_{0l}^{(\tiny{R})}(0)$ is determined
		by eqs. (\ref{E:rabi}), (\ref{E:four2}),
		describing the three-photon $l=3$ oscillations
		between the ground subbands $N=0$.
	}
	\label{fig1}
\end{figure}

\begin{figure}[h]
	\centering
	\includegraphics[width=1\columnwidth]{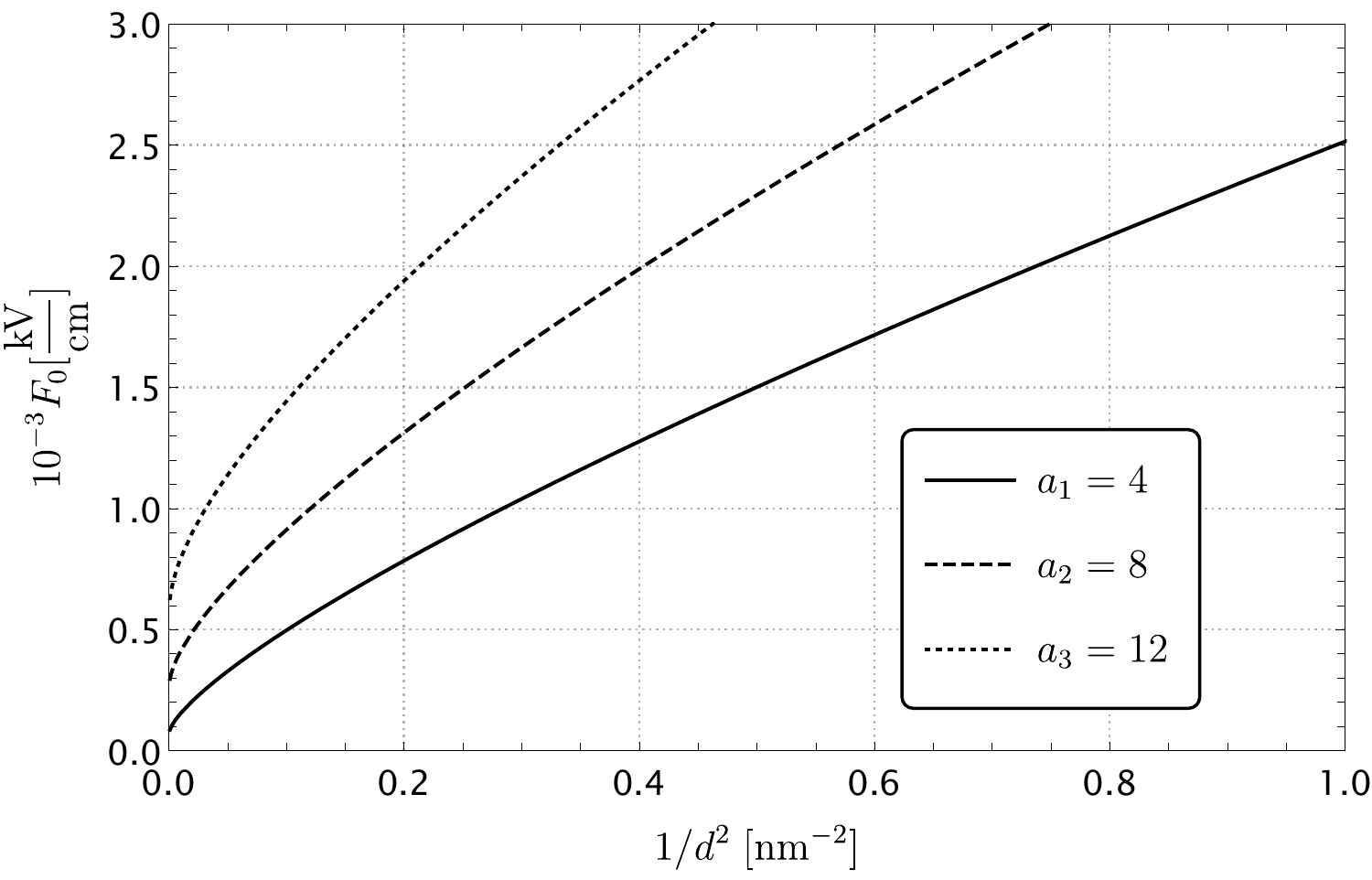}
	\caption{The isofrequency curves
		$\Omega_{03}^{(\tiny{R})}(0;d,F_{0})=a_j,~j=1,2,3$,
		linking the ribbon width $d$
		and electric field $F_{0}$.
		Eqs. ((\ref{E:four2}) and \ref{E:rabi})
		for the ground $N=0$ three-photon $l=3$
		transitions are employed.}
	\label{fig2}
\end{figure}

Eq. (\ref{E:probab}) describes periodic
oscillations with the Rabi frequency
$\Omega_{Nl}^{\left( R \right)} = 2\lambda_l$.
Under the condition $\omega_l/2 \ll |A_l |$
the transition probability
$w_N^{\tiny (l)} = \sin ^2 |A_l| t$ oscillates with the
Rabi frequency

\begin{equation}\label{E:rabi}
\Omega_{Nl}^{\left( R \right)}(k) = 2 |A_l(k)|.
\end{equation}
For the cases of the tunneling
$\gamma_N \ll 1,\Omega_{Nl}^{\left( R \right)}\equiv
\Omega_{N\tiny{tun}}^{\left( R \right)}
~\mbox{and the multiphoton}~\gamma_N \gg 1
$
regimes the Rabi frequency, determined
from eqs. (\ref{E:rabi}), (\ref{E:four2}) for the
zero longitudinal momentum $k$, reads

\begin{eqnarray}\label{E:rabilim}
\Omega_N^{\left( R \right)}(0) &&=
\frac{2}{3}\omega
\left\{
\begin{array}{cl}
\frac{1}{3}\exp\left(-\frac{\pi F_{\tiny{c}}^{(N)}}{2 F_{0}}
\right)~;\, &\gamma_N \ll 1
\\
\exp\left(l\right)\left( 4\gamma_N \right)^{-l}\sin^2 \frac{l\pi}{2}~;\,
&\gamma_N\gg 1
\end{array}
\right. \\
\label{E:crit}
F_{\tiny{c}}^{(N)} &&=\frac{\Delta_N^2}{4\hbar v_{\tiny{F}}e}, \nonumber
\end{eqnarray}
where $F_{\tiny{c}}^{(N)}$ is the breakdown electric field,
delimiting the active and suppressed tunneling
for the fields $F_{0} > F_{\tiny{c}}^{(N)}~\mbox{and}~
F_{0} < F_{\tiny{c}}^{(N)} $, respectively.
In the presence of the critical field
the electron in the AGNR with the effective mass $m=\Delta_N/2v_{\tiny{F}}^2$
acquires for the Compton wavelength
an energy comparable
to the energy gap $\Delta_N$.
Eqs. (\ref{E:rabi}), (\ref{E:four2}), and (\ref{E:rabilim})
allow us to trace the dependencies of the Rabi frequency
on the electric field and ribbon width in the
vicinity of the resonance $\omega_l \simeq 0$.
The photon assisted transitions are allowed only for the
odd numbers $l=1,3,5,\ldots$.
With increasing electric field $F_{0}$,
driving frequency $\omega = \frac{\mathscr{E}_N}{\hbar l}$ and
ribbon width $d$ the Rabi frequency $\Omega_{Nl}^{\left( R \right)}(0)$
increases for any regime. The Rabi frequency
$\Omega_{Nl}^{\left( R \right)}(0)$ according to eqs.
(\ref{E:rabi}), (\ref{E:four2}) as a function
of the ribbon width and electric field
is depicted in \cref{fig1}. \cref{fig2} shows the isofrequency
lines $\Omega_N^{\left( R \right)}(0;F_{0}, d)=\mbox{const}.$

\subsection{Intersubband transitions}

\subsubsection{Multiphoton assisted transitions}

If the detuning $\omega_l$ dominates
in eqs. (\ref{E:sol}), (\ref{E:probab}),
then, in view of the general results of time-dependent
perturbation theory \cite{LandLif}, the total probability
of the $l$-photon intersubband transition
per unit length per unit time
acquires the form

\begin{equation}\label{E:probab1}
W_N^{(\tiny{l})} = \frac{1}{2\pi}\int dk 2\pi \hbar  |A_l (k)|^2
\delta \left(\mathscr{E}_N (k) -l\hbar \omega \right),
\end{equation}
where the $\delta$-function reflects the energy
conservation in the system of the e-h pair and
absorbed $l$ photons. Substituting the coefficient
$A_l$ and energy $\mathscr{E}_N$
from eqs. (\ref{E:four2}) and  (\ref{E:enex}), respectively,
we obtain for the length density of the e-h pp rate

\begin{equation}\label{E:probab2}
W_N^{(\tiny{l})} = \frac{\pi^{\frac{1}{2}}}{36}
\frac{ \omega^2}{v_{\tiny{F}}}I_l (s_N)
G_l^{-\frac{1}{2}}(\omega)\sin^4 \frac{l\pi}{2},
\end{equation}
which in turn determines the $l$-photon absorption
coefficient $\alpha^{(l)}$ in eq. (\ref{E:coeff}).
In eq. (\ref{E:probab2}) the functions

\begin{equation}\label{E:intens}
I_l(s_N)=s_N^{-\frac{1}{2}}E^{-\frac{1}{2}}
\left(\sqrt{1- s_N^2}\right)
\exp\biggl\{ -\frac{2l}{s_N}\left[K(s_N)- E(s_N)\right]\biggr \}
\end{equation}
and

\begin{equation}\label{E:peak}
G_l (\omega ) = \frac{l\hbar \omega}{\Delta_N}-
\frac{2}{\pi s_N}E\left(\sqrt{1- s_N^2}  \right)
\end{equation}
are responsible for the spectral intensity and position of the
absorption singularity $\sim G_l^{-\frac{1}{2}}$, respectively.

For the limiting case $\gamma_N \gg 1$ the functions
$I_l(s_N)~\mbox{and}~G_l (\omega )$ become

\begin{equation}\label{E:intens1}
I_l(\gamma_N)=\left(\frac{2}{\pi}\right)^{\frac{1}{2}}
\exp \left( 2l\right) \left(  \frac{1}{16\gamma_N^2}\right)^l
\end{equation}
and

\begin{equation}\label{E:peak1}
G_l (\omega ) = \frac{l\hbar \omega}{\Delta_N}-
\left(1 +\frac{1}{4\gamma_N^2}  \right),
\end{equation}
respectively.

\begin{figure}[h]
	\centering
	\includegraphics[width=1.0\columnwidth]{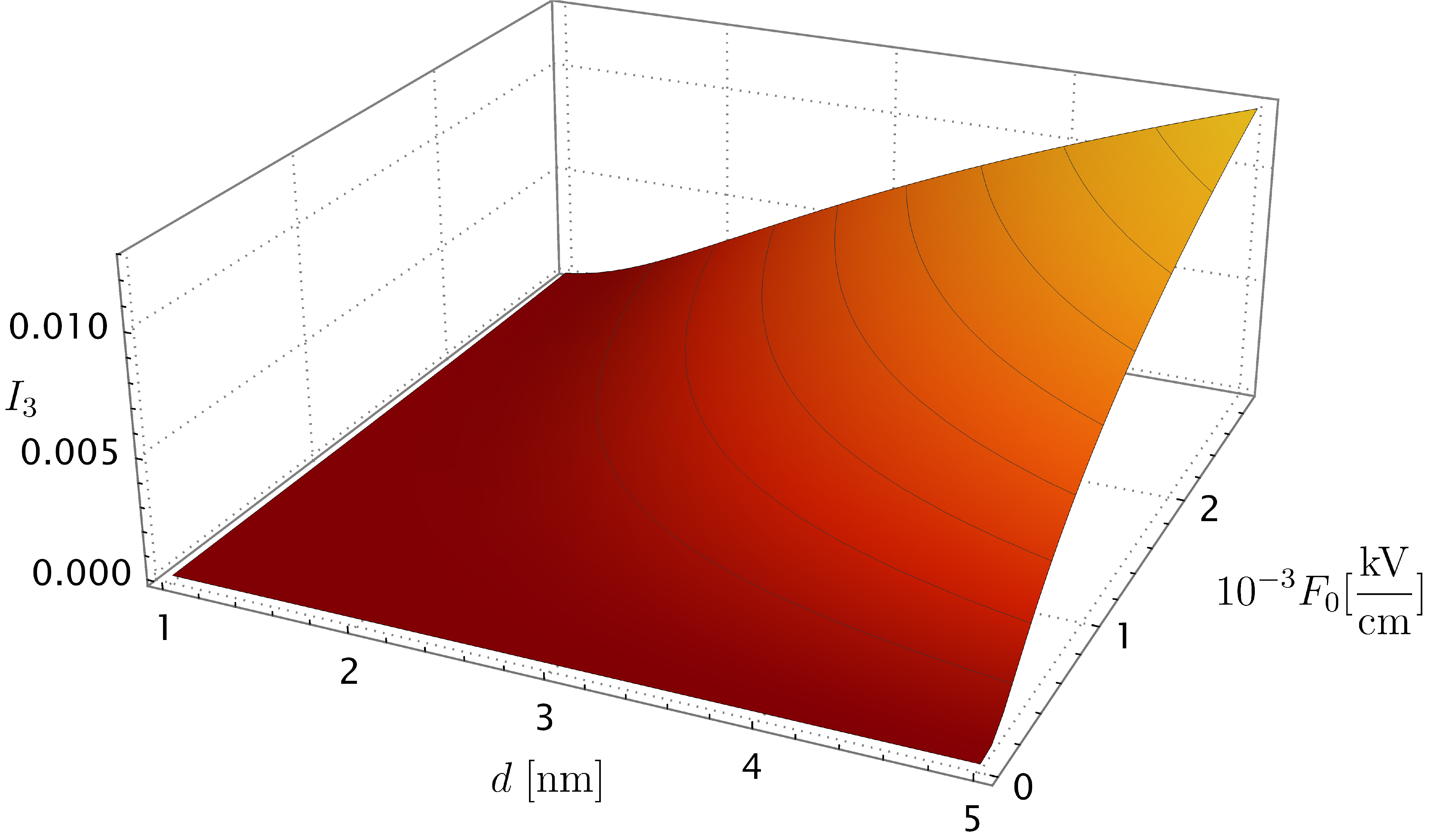}
	\caption{The dependence of the dimensionless
		peak intensity $I_3$ on the ribbon width $d$
		and electric field $F_{0}$.
		It is obtained from eq. (\ref{E:intens}) for $I_3$,
		adapted to the three-photon $l=3$ absorption,
		induced by the transitions
		between the ground $N=0$ subbands.}
	\label{fig3}
\end{figure}

\begin{figure}[h]
	\centering
	\includegraphics[width=1.0\columnwidth]{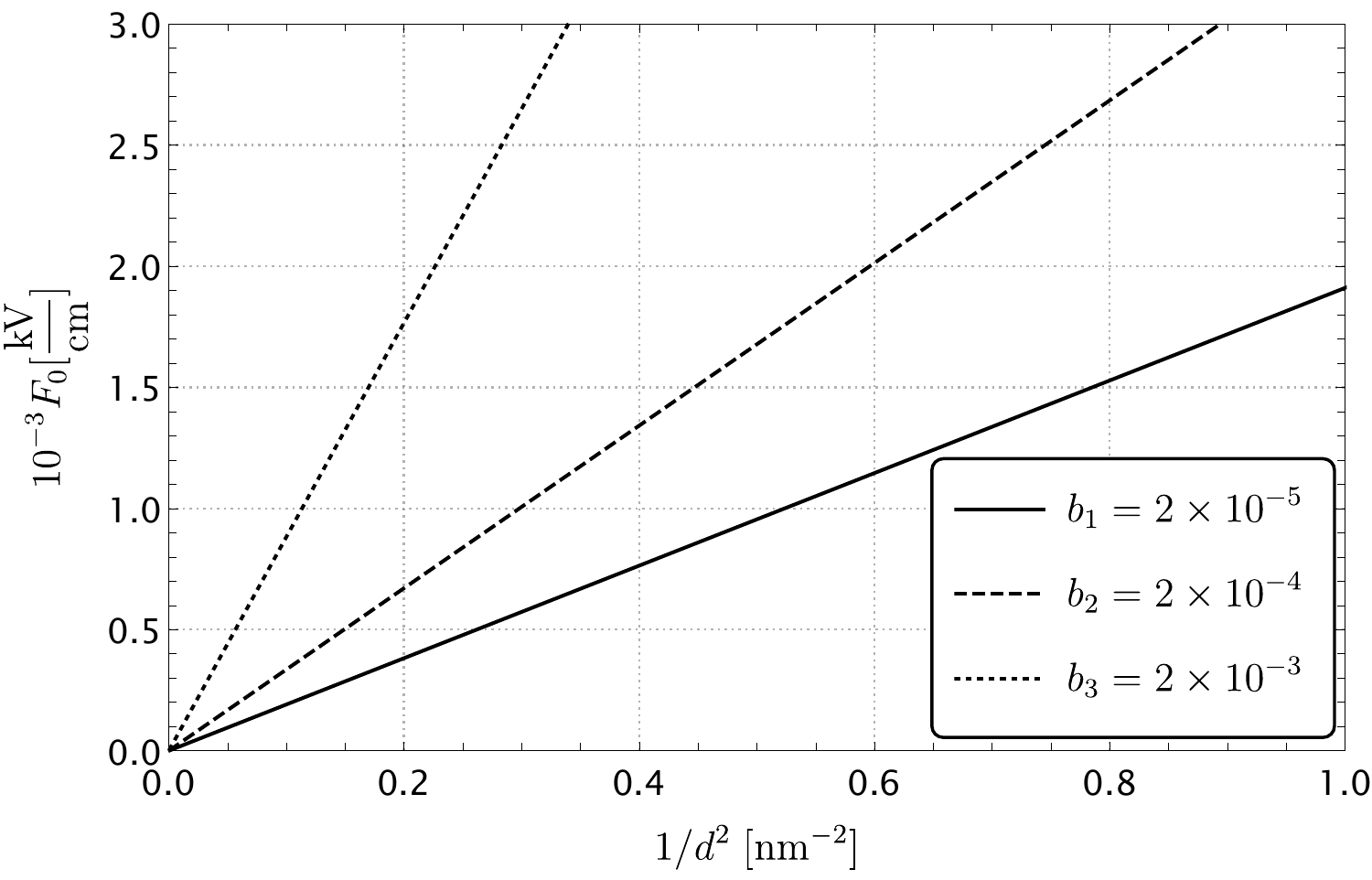}
	\caption{The isointensity curves
		$I_3(d,F_{0})=b_j,~j=1,2,3,$
		related to the ribbon width $d$
		and electric field $F_{0}$.
		Eq. (\ref{E:intens}) for $I_3$ is employed for the
		three-photon $l=3$ transitions between the ground
		$N=0$ subbands.}
	\label{fig4}
\end{figure}

Eqs. (\ref{E:probab2}), (\ref{E:intens1}) and (\ref{E:peak1})
explicitly demonstrate the dependencies of the multiphoton
intersubband absorption spectrum (\ref{E:coeff})
on the parity of the absorbed photon number $l$,
subband number $N$, ribbon width $d$ and electric
field $F_{0}$. As a result of the
saddle points $\varphi_{0,\pm1}$ interference
in the integrand of eq. (\ref{E:four1})
the spectral singularities $\sim G_l^{-\frac{1}{2}}~\mbox{in}~W_N^{(l)}$\,
(\ref{E:probab2}) are allowed only for the
odd numbers $l= 1,3,\ldots .$
The greater the photon number $l$ is, the less the
corresponding intensity $I_l$ in eq. (\ref{E:intens1}).
With increasing subband number $N$ the peak positions,
determined by the condition $G_l (\omega ) = 0$
in eq. (\ref{E:peak1}),
shift towards high frequencies and decrease in
intensity. As the ribbon becomes narrower, the peaks
move to the high frequency region $l\hbar\omega \sim d^{-1}$
and reduce in magnitude $\sim d^{{2l}}$.
The larger the electric field
$F_{0}$ is, the larger are both the
shift of the peak position
to higher frequencies $\sim F_{0}^2$,
and its maximum
$I_l\sim F_{0}^{2l}$. For the general case
for an arbitrary $\gamma_N$
the dependency of the peak intensity
$I_l$ (\ref{E:intens}) on the electric field
$F_{0}$ and ribbon width
$d$ are shown in \cref{fig3}. \cref{fig4}
demonstrates the isointensity
curves $I_l(F_{0},d)=\mbox{const.}$

\subsubsection{Tunneling}

In the case of small frequencies $\omega$
and considerable electric fields $F_{0}$
i.e. $\gamma_N \ll 1~ (\omega \rightarrow 0)$
the transitions happen due to the intersubband
tunneling in a practically stationary electric field
$F_{0}$. Based on the calculation
of the probability $W_N = \sum_l W_N^{(\tiny{l})} $
by means of $\sum_l$ by
$\int \frac{1}{\omega}d (l\omega)$  and eq.
(\ref{E:probab2}) by $W_N^{(\tiny{l})}$,
correctly reproduces the exponential behaviour

$$
W_N \sim \exp \left(-\frac{\pi F_{\tiny{c}}^{(N)}}
{ F_{0} }  \right),
$$
coinciding with those, obtained earlier
in Refs. \cite{Keld58, Keld65}.
The dimensionless prefactor
$\hbar v_{\tiny{F}} eF_{0}/\Delta_N^2$
appears to be incorrect. As pointed out
in Ref. \cite{Keld65}, this stems from the fact
that the limiting transition $\omega \rightarrow 0$
applies to the results based on the averaging
over the finite period $T=2\pi/\omega$ of the
oscillating electric field. The corresponding
correct analysis
of the intersubband tunneling
should however treat the electric field to be
adiabatically slow from the beginning.

The proper determination
of the differential probability
$w_N (\vec{k})$ started in the 1950s
on account of the vacuum decay in the presence
of a constant electric field \cite{Schwin}.
This study was continued in the 70s further.
The various approaches,
namely the imaginary time method
\cite{Pop1}, parabolic cylinder functions
\cite{Nik}, \cite{Nik1}, \cite{Pop}, \cite{KlimM},
\cite{GavGit}, \cite{KimP},
Riccati \cite{PopMar}, \cite{Pop}
and quantum kinetic
equations (QKE) were proposed
\cite{Fedot}, \cite{Heben}, \cite{Schm}, \cite{Akal14}.
Notice, that the probability $W_N$ can be calculated within
the present approach.
Indeed, it was shown that the set of initial eqs. (\ref{E:set1})
is trivially reduced to the oscillator-like
equations for the parabolic cylinder functions
\cite{KlimM}, \cite{GavGit}, while the set (\ref{E:set2})
for the functions $f_{1,2}(k,t)$ is
equivalent to the Riccati \cite{Pop}, \cite{PopMar}
and/or (QKE)
\cite{Heben},\cite{Schm},\cite{Akal14}
equations for the
functions $|f_1(k,t)|$ and $|f_1(k,t)|^2$, respectively.
The details of the transformation of eqs. (\ref{E:set2})
into the QKE are given, in particular, in Ref. \cite{Akal14}.
In addition, Fedotov \emph{et al} \cite{Fedot},
based on the exact solution to QKE,
have calculated the differential distribution
function for the interband transition probability,
i.e., the mean number of pairs $w_N (\vec{k})$ created
in a given quantum state,
completely coinciding with those, derived by others
of the above-listed methods.
The underlying analysis is transparently presented in
Refs. \cite{Akal14,Fedot} and we provide here for
reasons of brevity only their results
in an explicit form

\begin{eqnarray}\label{E:wW}
\begin{array}{l}
w_N(\vec{k})=2\exp\left[ -
\frac{\pi (\varepsilon_N^2 + \hbar^2 v_{\tiny{F}}^2 \vec{k}_{\perp}^2
    )}{\hbar v_{\tiny{F}} e F_{0}} \right];\\
W_N^{(n)}= \frac{1}{(2\pi)^n T}\int w_N(\vec{k})d^n \vec{k},
\end{array}
\end{eqnarray}
where $W_N^{(n)}$
is the spatial
density of the tunneling probability rate
\cite{All} in the $n$-dimensional structure.
In eq. (\ref{E:wW}) the prefactor two takes into account
the spin projections,
$\vec{k}[\vec{k}_{\perp},k  ]$ is the wave vector
and $\varepsilon_N$ is the energy gap.
Integration in eq. (\ref{E:wW}) over $k$ implies
$\int dk = e F_{0} T/\hbar $ \cite{KlimM},
where $T$ is the total (infinitely large)
lifetime of the dc electric field
(see Refs. \cite{Nik1, GavGit} for details).

For the quasi-1D AGNR $(n=1)$ eq. (\ref{E:wW}),
in view of the two valleys $\left(K^{\pm} \right)$
\cite{KlimM, Fill1, Akal16}, generates the mean total number of the e-h pairs
per unit length per unit time, created due to the
tunneling transition
via the subband gap $\Delta_N = 2\varepsilon_N
$ (eq. (\ref{E:subb})),

\begin{equation}\label{E:probrib}
W_N^{(1)}\equiv W_{N\tiny{{tun}}} = \frac{2e F_{0}}{\pi\hbar}
\exp\left[ -
\frac{\pi F_{\tiny{c}}^{(N)}}{ F_{0}} \right].
\end{equation}
The same result, accurate to the spin and valley
factors, has been derived by Gavrilov and Gitman
\cite{GavGit} for the
1D spatial states
in the framework of an exact solution to the Dirac equation.
In addition, just the same approach has been employed
to study the mathematically exactly solvable problem of particle
production from a QED vacuum by the Sauter-like
and peak time depending electric field \cite{AdorGG}.
The total length density of the e-h pp rate
$W_N = \sum_N W_{N\tiny{{tun}}}$ is derived
from eq. (\ref{E:probrib}). In the limiting case
$d\rightarrow \infty$ the latter equation with the
replacement $\sum_N~\mbox{by}~d\int d \bigl(\frac{N}{d}\bigl)$
results, as expected, in the square density rate
for the gapless $(d^{-1}\sim \Delta_N \rightarrow 0)$
graphene layer

$$
W_{\tiny{g}}= \frac{1}{\pi^2 v_{\tiny{F}}^{\frac{1}{2}}}
\biggl( \frac{eF_{0}}{\hbar}  \biggr)^{\frac{3}{2}},
$$
presented, in particular, in Refs. \cite{KlimM, Akal16}.

A doubled rate $2W_{v}$ of the electron-positron
pair production from the "1D vacuum" can be obtained
from eq. (\ref{E:probrib}) by replacing
$v_{\tiny{F}}~\mbox{by}~c~\mbox{and}~\Delta_N~\mbox{by}~2mc^2.$
Eq. (\ref{E:wW})
has been used by other authors; for the 3D and 2D spaces
eq. (\ref{E:wW}) reproduces the vacuum
$(\varepsilon = mc^2, v_{\tiny{F}}=c)$
\cite{Nik1, Schwin, All} and gapless
$(\varepsilon = 0)$ \cite{KlimM} graphene rates,
respectively. The rate (\ref{E:probrib}) differs from
the one calculated in Ref.
\cite{Akal14} by a factor of four, because
of the spin and valley factors which both equal two.
Note that eq. (\ref{E:wW}), describing the time-independent
effect, can be obtained by the WKB method, while
for the multiphoton assisted processes
(see eq.(\ref{E:intens1})) the semiclassical
approximation is inappropriate \cite{Taya21}.

\begin{figure}[h]
	\centering
	\includegraphics[width=1.0\columnwidth]{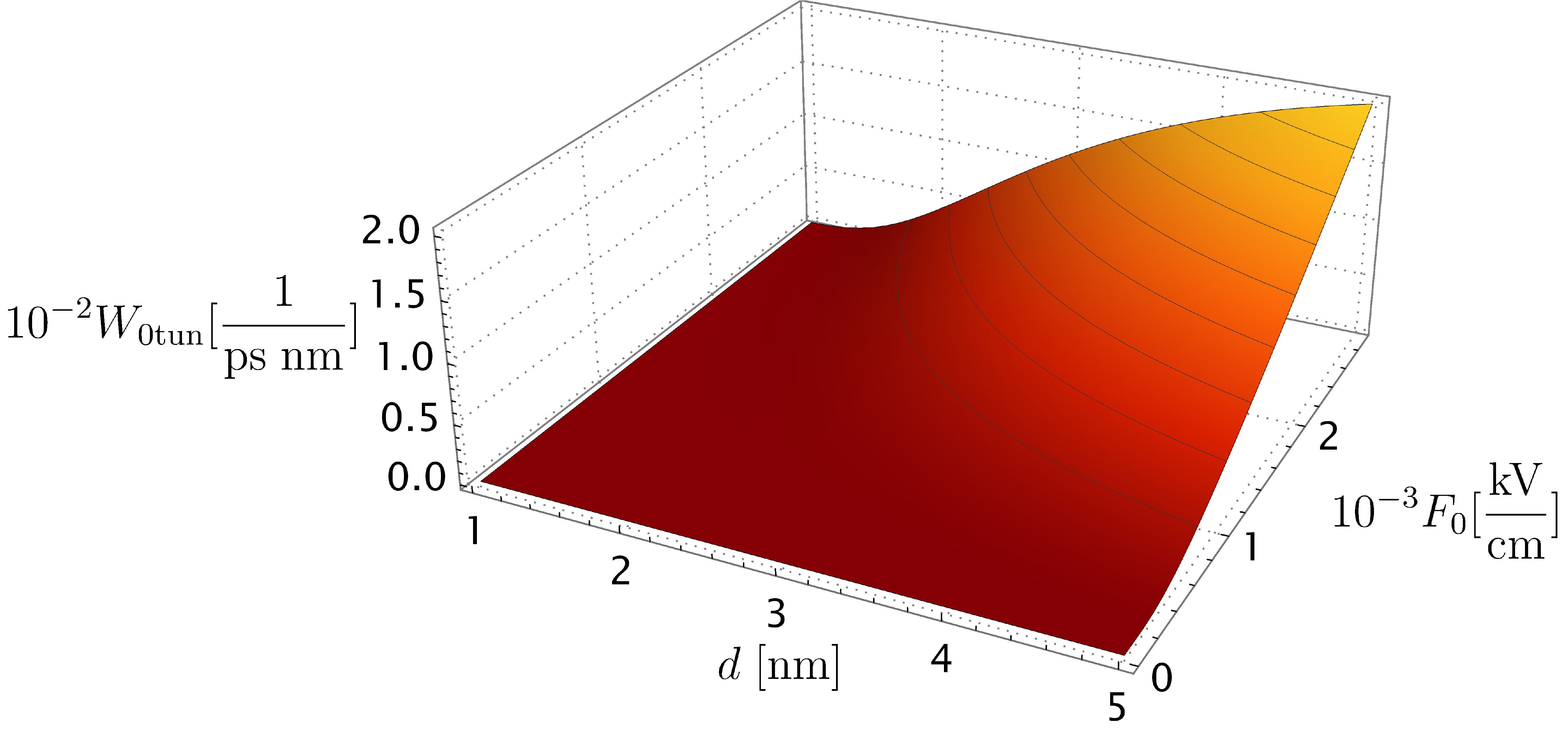}
	\caption{The length density of the
		pair production rate $W_{\tiny{0{tun}}}$ (\ref{E:probrib}),
		caused by the electron tunneling
		between the ground $N=0$ subbands
		in the ribbon with width $d$ in the presence
		of an electric field $F_{0}$.}
	\label{fig5}
\end{figure}

\begin{figure}[h]
	\centering
	\includegraphics[width=1.0\columnwidth]{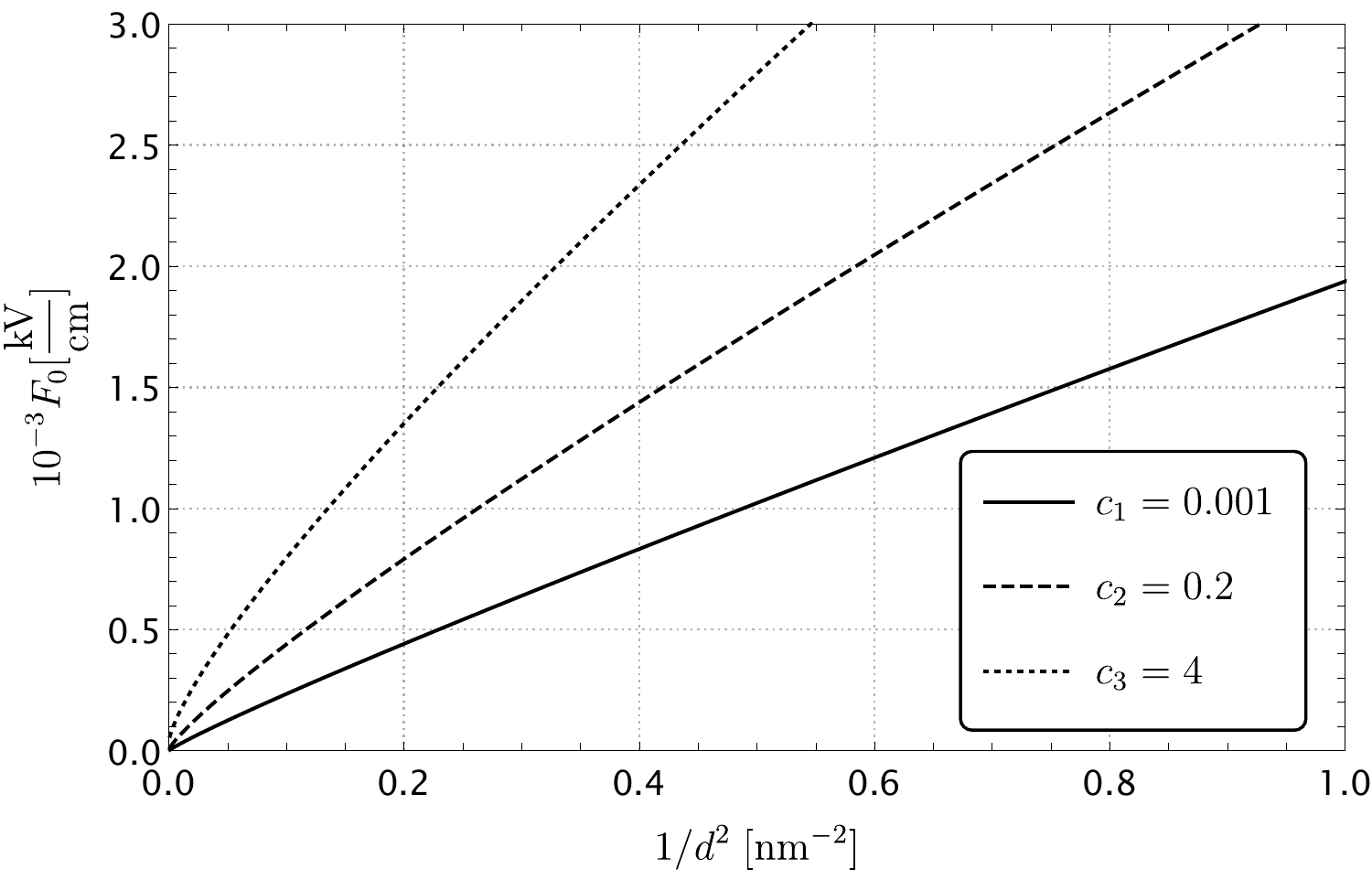}
	\caption{The isorate diagrams
		$W_{\tiny{0{tun}}}(d,F_{0})~(\mbox{nm ps})^{-1}=c_j,~j=1,2,3.$
		calculated from eq. (\ref{E:probrib}) for the
		tunneling between the ground $N=0$ subbands.}
	\label{fig6}
\end{figure}

Clearly, the length density
of the e-h pp tunneling
rate $W_{N\tiny{{tun}}}$ (\ref{E:probrib})
increases with both increasing the electric
field $F_{0}$ and ribbon width $d$
according to \cref{fig5}. The isorate diagrams
$W_{0\tiny{{tun}}}(F_{0},d)= \mbox{const}$
are depicted in \cref{fig6}. It is appropriate
here to point out the common property of the
isovalues diagrams. The Figs. 2 and 6
demonstrate the practically linear
relationship v.r.t. the considerable
electric fields $F_{0}$ and the
square of the reciprocal width $1/d^2$ of the
narrow ribbons. As expected, deviations
from the linear law occur for weak
electric fields and wide ribbons.
The reason for this is that the latter
closely resemble more the graphene
layer than a ribbon. \cref{fig4} shows the strict
linear dependence $F_{0}\sim 1/d^2$.
Note here that the correct description
of the wide ribbon and the graphene layer implies the summation
over the subband index $N$ of the rates
$w_N^{(l)}(k) (\ref{E:probab}),(\ref{E:rabi}), W_N^{(l)} (\ref{E:probab2})
~\mbox{and}~W_{N tun}$ (\ref{E:probrib}) for the
Rabi oscillations, multiphoton assisted and
tunneling transitions, respectively.

\begin{figure}[h]
	\centering
	\includegraphics[width=1.0\columnwidth]{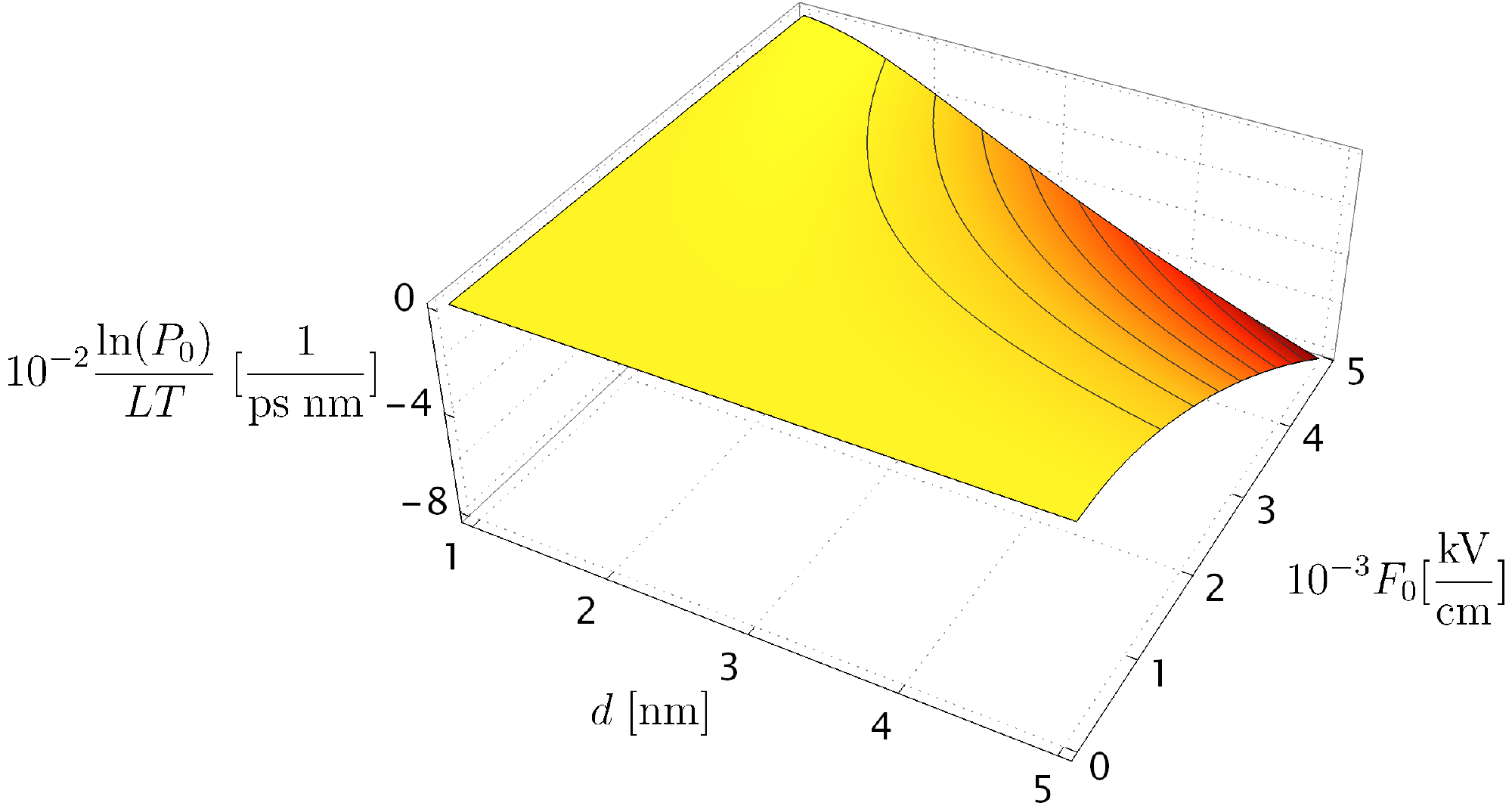}
	\caption{The logarithm of the length density rate of
		the probability $\mathscr{P}_0$
		as a function of the electric
		field $F_{0}$ and ribbon width $d$,
		calculated from eq. (\ref{E:remain}) for the
		ground transition $N=0$. $T~\mbox{and}~L$
		are the radiation exposition time and
		ribbon width $d$, respectively.}
	\label{fig7}
\end{figure}

Using the relation between the differential
probability $w_N(\vec{k})$ (\ref{E:wW})
and an AGNR to an AGNR $N$ transition probability
$\mathscr{P}_N$ \cite{GavGit,AdorGG},
i.e., the probability for an AGNR to
electronically remain an AGNR
on account of the tunneling between the $N$ subbands

$$
\mathscr{P}_N = \exp\biggl\{\int d\vec{k}
\ln [ 1 - 2 w_N(\vec{k})] \biggr\}
$$
and integrating over the momentum $k$
by the same method, taken in eq. (\ref{E:wW})
we obtain

\begin{equation}\label{E:remain}
\frac{\ln \mathscr{P}_N }{LT}=\frac{2e F_{0}}{\pi\hbar}
\ln \biggl[1  - \exp\biggl(-\frac{\pi F_{\tiny{c}}^{(N)}}{F_{0}} \biggr)\biggr].
\end{equation}
The greater the electric field
and the wider the ribbon are,
the less is the ribbon stability $\mathscr{P}_N$.
The dependence of the probability
$\mathscr{P}_0$ w.r.t. the ground
transition $N=0$ on the electric field $F_{0}$
and width $d$ is shown in  \cref{fig7}.

\begin{figure}[h]
	\centering
	\includegraphics[width=1.0\columnwidth]{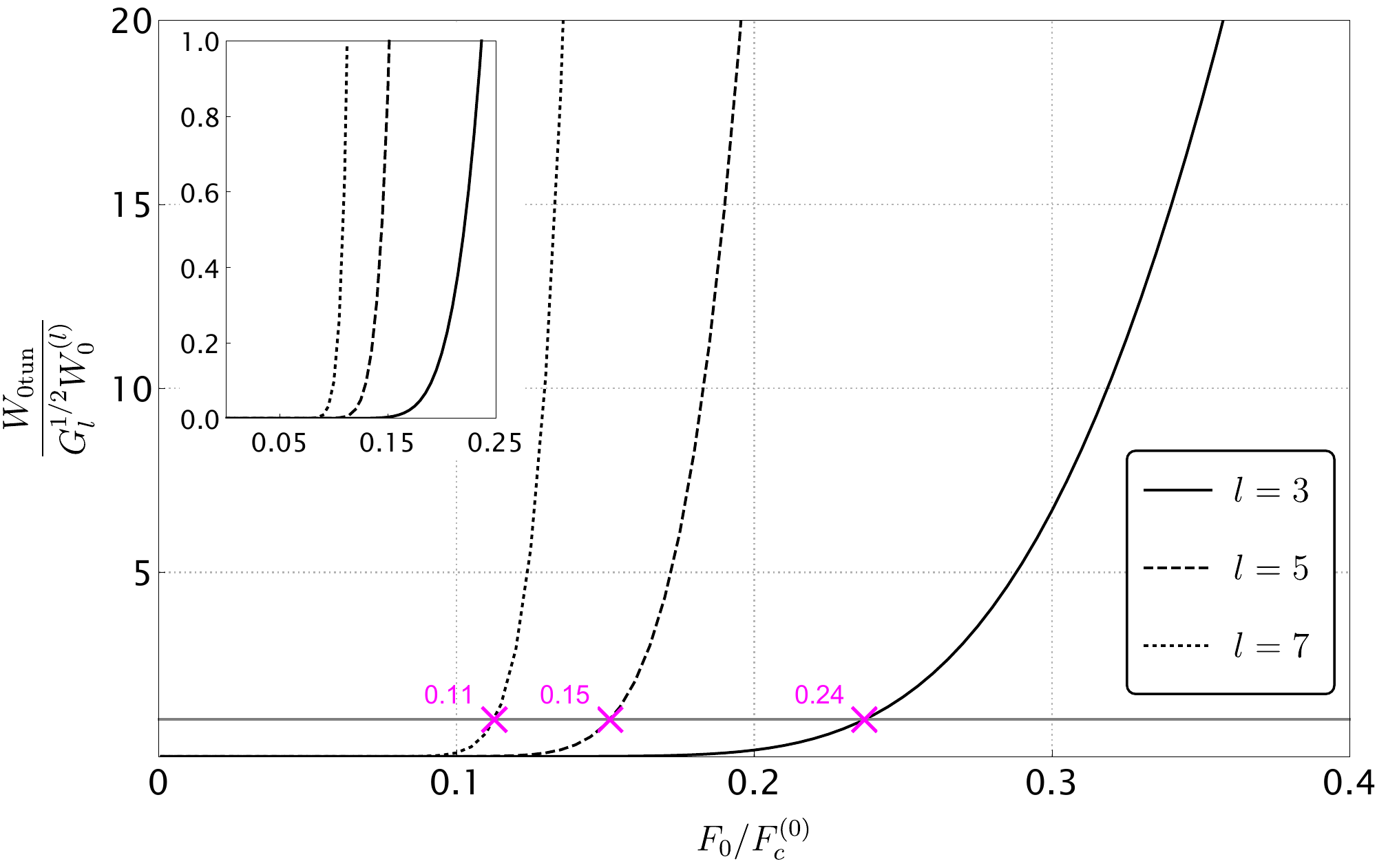}
	\caption{The ratio of the tunneling  $W_{\tiny{0{tun}}}$
		(\ref{E:probrib}) and photon assisted
		$G_l^{\frac{1}{2}}W_{0}^{(l)}$
		(\ref{E:probab2}) length density rates
		as a function of the dimensionless electric field
		$F_{0}/F_{c}^{(0)}$
		for the $l$-photon $(l=3,5,7)$ transitions
		between the ground $N=0$ subbands for which
		$F_{c}^{(0)}=2.0\cdot10^3~\mbox{kV/cm}$.}
	\label{fig8}
\end{figure}

In order to highlight the contribution of the
time-dependence of the electric field
to the e-h pair production we investigate the dependence
of the ratio of the tunneling  $W_{0\tiny{{tun}}}$
(\ref{E:probrib}) and photon assisted
$G_l^{\frac{1}{2}}W_{0}^{(l)}$
(\ref{E:probab2}) rates on the electric field
$F_{0}$.
\cref{fig8} shows this ratio as a function of
the electric field $F_{0}$. These graphs,
based on eqs. (\ref{E:probab2}) and (\ref{E:probrib}),
demonstrate that
in the region of weak electric fields
$F_{0}/F_{c}^{(0)}\ll 1$
the photon assisted rate surpasses that of tunneling.
With the electric field approaching the values
$F_{0}/F_{c}^{(0)} \simeq 1$
the advantage of the
photon assisted mechanism reduces, and both rates become
comparable and further, for
$F_{0} \geq F_{c}^{(0)}$,
the tunneling process
dominates that of photonic absorption.

The reason for the above described evolution is that
the electric field change generates
a change of the transition regime.
For weak electric fields
$F_{0}\ll F_{c}^{(0)}$
and low frequencies $\omega< \Delta_0 / \hbar$
the intersubband tunneling time
$\tau_0= \Delta_0/eF_{0}v_{\tiny{F}}$
exceeds the half-cycle $ T/2 = \pi/\omega,~
(\tau_0 > T/2,~\gamma_0 \gg 1)$
and the fast oscillating $(\omega > \tau_0^{-1})$
field prevents the tunneling and
promotes the involvement of the
$l \simeq \Delta_0/ \hbar \omega$ photons
in the intersubband transitions.
With the growth of the electric field towards
values exceeding the critical one
$F_{0} \geq  F_{c}^{(0)},$
the opposite conditions $\tau_0 <T/2,~\gamma_0 \geq 1$
allow us to treat the intersubband transitions as
a tunneling process induced by the nearly constant
$(\omega < \tau_0^{-1})$ electric field $F(t)= F_{0}$.
The role of the time dependence of
a relatively weak
$F_{0}< F_{c}^{(0)}$
electric field for the enhancement
of the pp rate, revealed here for the AGNR, is absolutely
analogous to that for the electrically biased graphene layer,
studied numerically in Ref. \cite{Akal16}
(see also \cite{Avet,Fill1} for details). However,
we refrain from a detailed
quantitative comparison of our results to those
presented for the graphene layer. The reason for this
is that the prefactor in eq. (\ref{E:probrib})
$\sim F_{0}$ and the states energy density
factor in eq. (\ref{E:probab2})
$\sim G_l^{-\frac{1}{2}}$ differ from
the corresponding ones for the 2D structures,
namely $\sim F_{0}^{\frac{3}{2}}$
\cite{Akal16} and $\sim \mbox{const}$, respectively. Clearly,
accounting for the electron interaction with
the phonons and impurities,
manifesting itself in replacing the root singularity in eq.
(\ref{E:probab2}) by a finite value
would lead to a more adequate description
of the electronic and optical effects in AGNR.
However, this problem is a subject of further
possible consideration.

\cref{fig8} allows us to reveal the relationship between
the key electric fields, characterising the given
$N$ intersubband transition. First one is
the threshold field
$F_{0thr}^{(l)}$,
determined by the critical resonant
Keldysh parameter $\gamma_N^{(l)}=1,$
in which (see eq. (\ref{E:four1}))
the frequency $\omega$ satisfies the
resonant condition $\mathscr{E}_N (0)= l\hbar \omega$,
where $\mathscr{E}_N (0)$ is determined
from eq. (\ref{E:enex}).
The field $F_{0thr}^{(l)}$
qualitatively delimits the tunneling
$(F_{0}< F_{0thr}^{(l)},
\gamma_N^{(l)} < 1)$
and $l$-photon assisted
$(F_{0}> F_{0thr}^{(l)},
\gamma_N^{(l)} > 1)$
type transitions. On account of
the elliptic integral of the second kind,
$E[(1 +\gamma_N^2 )^{-\frac{1}{2}}]$
it changes smoothly in the vicinity of $\gamma_N \simeq 1.$
and we set $E(2^{-\frac{1}{2}})= 1.23$
to obtain the resonant Keldysh parameter
$\gamma_N^{(l)}$ (\ref{E:four1})

$$
\gamma_N^{(l)2}= \frac{1}{2}+
\sqrt{\frac{a^2}{4} + a};
~a= \left(\frac{\pi}{2lp_N}\right)^2;~p_N
=\frac{F_{0}}{F_{\tiny{c}}^{(N)}}.
$$
This equation links the resonant Keldysh parameter
$\gamma_N^{(l)}$ with the electric field $F_{0}$
and number of photons $l$ in the intermediate region
$\gamma_N^{(l)}\simeq \leq 1$.
For $\gamma_N^{(l)}=1$ the exact value of the
threshold dimensionless electric field, calculated
from the resonant condition, reads

$$
p_{\tiny{N}thr}^{(l)}= \frac{\pi}{\sqrt{2} l}.
$$
The second key, so-called balanced
electric field
$F_{0b}^{(l)}$,
equalizes the tunneling and $l$-photon
assisted rates.
Calculating the threshold
$p_{\tiny{N}thr}^{(l)}= \frac{\pi}{\sqrt{2} l}$
and balanced $p_{\tiny{N}b}^{(l)}$
dimensionless electric fields for the ground transition
$N=0$ from the given above equation and from \cref{fig8},
respectively, we present the result of their
comparison in \cref{tab:tab1}.

\begin{table}[h]
     \centering
     \scalebox{1.2}{
     \begin{tabular}{ |c|c|c|c| }
         \hline
         l & $p^{(l)}_{0thr}$ & $p^{(l)}_{0b}$ &
$p^{(l)}_{0thr}$/$p^{(l)}_{0b}$ \\ \hline
         3 & 0.74             & 0.24           & 3.0 \\ \hline
         5 & 0.45             & 0.15           & 3.0 \\ \hline
         7 & 0.32             & 0.11           & 2.9 \\ \hline
     \end{tabular}}
     \caption{The dimensionless threshold $p_{0thr}^{(l)}=
F_{0thr}^{(l)}/ F_{c}^{(0)}$ and
balanced
$p_{0b}^{(l)}=F_{0b}^{(l)}/F_{c}^{(0)}$
electric fields, delimiting the $l$-photon assisted
and tunneling regimes $(\gamma_N^{(l)}=1)$ and providing the balance
between its rates $(W_{0^{(l)}}= W_{0tun})$,
respectively. Fields scaled to the critical electric field
$F_{c}^{(0)}$.}
\label{tab:tab1}
\end{table}
The ratio of the electric fields under discussion, being
$p_{0thr}^{(l)}/p_{0b}^{(l)}=3$,
does not depend on the number of photons $l$.
Thus, experimentally measuring the ratio of the rates
$W_{\tiny{N}}^{(l)}$ and $W_{Ntun}$
as a function of the electric
field, we find the balance field
$p_{\tiny{N}b}^{(l)}$
and in the case of known
$p_{\tiny{N}thr}^{(l_0)}$
for the specific $l_0$,
physically important dimensionless
electric fields for other values of $l$
can be found.

The effect of the oscillating character
of the electric field on the Rabi oscillations
is qualitatively the same as that
on the above discussed intersubband
transitions. For weak electric fields
$F_{0}\ll F_{\tiny{c}}^{(N)}$
their time periodic resonant oscillations
significantly increase  the
multiphoton assisted Rabi frequencies
$\Omega_{\tiny{Nl}}^{(R)}$,
making them much greater than
the corresponding $\Omega_{\tiny{N}tun}^{(R)}$,
induced by the tunneling
for approximately constant electric field.
With increasing electric field strength
towards the values
$F_{0}\simeq F_{\tiny{c}}^{(N)}$
these Rabi frequencies align
$(\Omega_{\tiny{Nl}}^{(R)}
\simeq\Omega_{\tiny{N}tun}^{(R)})$,
then the electric field time dependence
becomes ineffective
$(\Omega_{\tiny{Nl}}^{(R)}
< \Omega_{\tiny{N}tun}^{(R)})$.
The electric fields, providing the
Rabi frequency balance
$(\Omega_{\tiny{Nl}}^{(R)}
=\Omega_{\tiny{N}tun}^{(R)})$,
decrease with an increasing number
of involved photons $l$. The dependence
of the ratio
$\Omega_{\tiny{N}tun}^{(R)}
/\Omega_{\tiny{Nl}}^{(R)}$
on the dimensionless electric field
$F_{0}/F_{\tiny{c}}^{(N)}$
closely resembles that presented in \cref{fig8}.

All aforementioned conclusions in this section,
related to the ground $N=0$ e-h subband, apply qualitatively also
for excited ones with $N\neq 0$. It is reasonable to note here
that, similar to works \cite{All}, \cite{Akal19},
\cite{Akal16}, \cite{KimP}, \cite{Akal14},
our approach ignored
the collisions between the created pairs and backreaction
of their inherent electric field to the applied external one.
However, these effects might be expected
to be insignificant due to the relatively small
density of the created e-h pairs, that in turn depends
not only on the electric field magnitude $F_{0}$,
but on the exposure time $T$.
In any case these phenomena require
special consideration, in particular, in the framework of
a quantum kinetic equation \cite{Fedot}.

\subsection{Estimates of the expected experimental values}

Focussing on possible experiments, we estimate the expected
values for the gapped AGNR for a width of 2 nm
exposed to a light wave of $\omega = 330~\mbox{ps}^{-1} $
$(\lambda = 5.4~\mu\mbox{m})$ and electric field
$F_{0} = 500~\mbox{kV/cm}$.
This corresponds to a light intensity
$I=6.5\cdot 10^5~\mbox{kW/cm}^2$
and obeys the resonant condition $G_l = 0$ (\ref{E:peak})
for the number of photons $l=3$ and ground $(N=0)$
energy gap $\Delta_0 = 2\varepsilon_0 = 0.69~\mbox{eV} $
(\ref{E:subb}).
The multiphoton and tunneling rates
$W_{0}^{(\tiny{3})}\cdot
G_{\tiny{3}}^{\frac{1}{2}}~\mbox{and}~W_{0tun}$,
calculated from eqs. (\ref{E:probab2}) and (\ref{E:probrib}),
respectively, become
$0.86\cdot 10^{-4}~\mbox{1/nm ps}~\mbox{and}~1.4\cdot 10^{-4}~\mbox{1/nm ps}$,
respectively. For the electric field
$F_{0} = 470~\mbox{kV/cm}$ they reach a balance
equal to $1.4\cdot 10^{-3}~\mbox{1/nm ps}$
and with increasing the electric field
the tunneling
mechanism dominates that of multiphoton transitions.
The "to remain" probability $\mathscr{P}_N$,
calculated from eq. (\ref{E:remain}) for the
ground $N=0$ subbands and electric fields
$F_{0} = F_{c}^{(0)}=
2.0\cdot10^3~\mbox{kV/cm}$, reads
$(LT)^{-1}\ln \mathscr{P}_0 = - 8.48~ \mbox{1/nm ps}$.

The chosen ribbon $(d)$ and electric field $(F_0, \omega)$
characteristics result in
$\Omega_{03}^{(\tiny{R})}(0) \simeq 1.21~\mbox{ps}^{-1}$
(eqs. (\ref{E:rabi}), (\ref{E:four2}))
for the Rabi frequency and
$\Omega_{03}^{(\tiny{R})}(0)/
\omega \simeq 3.7\cdot10^{-3} \ll 1$
for the frequencies ratio.
Note that the vacuum related breakdown electric field
$F_{c}^{(v)}$ exceeds its counterpart
$F_{c}^{(0)}=2.0\cdot10^3~\mbox{kV/cm}$
for the ground gap $\Delta_0$ by a factor of $10^{10}$.

We believe that the
analytical approach developed here
contributes to gaining insights into
the physics of the intersubband transition in AGNR
and QED vacuum decay, both media being subject
to a strong light wave. Also, we hope the
estimates of the expected experimental values
could be useful for further studies of graphene nanoribbons
and their applications in opto- and microelectronics
as well as the vacuum phenomena.

\section{Summary and conclusion}\label{S:conc}

In summary, we have developed an analytical approach
to the problem of the Rabi oscillations and
intersubband absorption of a strong light wave in an armchair
graphene nanoribbon (AGNR). Based on the Dirac equation, describing
the massless electron in the vicinity of the Dirac
points, we have
derived analytical expressions for the length density
of the electron-hole differential pair production (pp) rate.
The resonant approximation, implying a balance
between the photons energies and intersubband
quasienergetic gaps,
has been employed.
This rate in turn determines explicitly
the Rabi oscillation frequency and
absorption coefficient for tunneling
and multiphoton assisted
intersubband transition regimes.
The obtained results allow us to trace
the explicit dependencies of the
Rabi frequency and pp rate on the ribbon
width, electric field strength and parity of the involved
photons. The odd-photon absorption spectra demonstrate the
reciprocal square root singularities
in the vicinity of the resonant frequencies. With
increasing the electric field and widening
the ribbon both the Rabi frequency and pp
rate increase. For relatively weak electric fields
the oscillating character of the electric field enhances
the intersubband transitions and multiphoton assisted
effects contribute significantly stronger than the tunneling ones.
With further increase of the electric field
these effects become equal and finally the tunneling mechanism
surpasses that of multiphoton processes.
The latter
dependence completely correlates with the
previously numerically calculated one for the gapped
graphene layer.
Estimates of the expected values show that the theoretically
predicted dependencies for the Rabi oscillations
and multiphoton absorption
spectra can be observed experimentally for realistic AGNR
subject to readily available light fields.
The results
presented above can be qualitatively extended to the
quantum electrodynamic vacuum decay in the presence of
strong time-oscillating electric fields
and the AGNR can be treated as a condensed
matter medium for the study of particle-antiparticle
creation processes, induced by the intensive time-dependent
electric fields.

\section{Acknowledgments}\label{S:Ackn}
The authors are grateful to S.~P.~Gavrilov for
many useful discussions and valuable comments
as well as M. Pyzh for significant assistance
in numerical calculations
and graphics.

\end{document}